\documentclass[12pt]{article} 
\usepackage{graphicx,amssymb,epsfig,rotate} 
\def\gsim{\lower0.5ex\hbox{$\:\buildrel >\over\sim\:$}}
\def\lsim{\lower0.5ex\hbox{$\:\buildrel <\over\sim\:$}}

\newcommand{\bea}{\begin{eqnarray}}
\newcommand{\eea}{\end{eqnarray}}

\begin{document}

\begin{center} 
{\Large \bf Resonant Leptogenesis with nonholomorphic\\
R-Parity violation and LHC Phenomenology}\\
\vskip 0.5in
{\bf Joydeep Chakrabortty$^\dagger$\footnote{E-mail: joydeep@hri.res.in}, 
Sourov Roy$^\ddagger$\footnote{E-mail: tpsr@iacs.res.in}\\}
\vskip 0.3in
{$^\dagger$ \sl Harish-Chandra Research Institute,
Chhatnag Road, Jhunsi, Allahabad  211 019, India.\\}
\vskip 0.1in
{$^\ddagger$ \sl Department of Theoretical Physics, Indian Association for the 
Cultivation of Science, 2A $\&$ 2B Raja S.C. Mullick Road, Kolkata 700 032, India.\\}
\end{center} 

\vskip 1 cm

\begin{abstract}
In R-parity violating supersymmetric models both leptogenesis and the correct neutrino masses 
are hard to achieve together. The presence of certain soft nonholomorphic R-parity violating terms 
helps to resolve this problem. We consider a scenario where the lightest and the second-lightest 
neutralino are nearly degenerate in mass and enough CP-asymmetry can be produced through resonant leptogenesis. 
In this model, the lighter chargino and the lightest neutralino are highly 
degenerate. We have relatively lighter gauginos which can be produced at the LHC leading to heavily ionizing charged 
tracks. At the same time this model can also generate the correct neutrino mass scale. Thus our scenario 
is phenomenologically rich and testable at colliders.
\end{abstract}

\pagebreak

\noindent
\section*{Introduction}
Baryon Asymmetry of the Universe (BAU) is one of the major challenges of cosmo-particle physics 
whose theory is not yet convincing. The observed baryon asymmetry is \cite{nakamura-pdg}
\bea
X_B \equiv \frac{n_B -n_{\bar B}}{\circledS} = \frac{n_B}{\circledS} \approx (7.2 - 9.2) \times 10^{-11}~~(95\% ~{\rm C.L.}), 
\eea
where $n_B$ is the number density of baryons, $n_{\bar B}$ is that of anti-baryons, and $\circledS$ is the entropy density. 
Leptogenesis \cite{leptogenesis}, leading to lepton asymmetry which partly gets converted 
into the baryon asymmetry through sphaleron processes is thought of as a good candidate to describe the 
matter-antimatter discrepancy of the universe i.e., BAU. Within the supersymmetric standard model (SSM), the final 
baryon asymmetry is related to the initial lepton asymmetry by $B = -\frac{32}{60}L$, where $B$ is the net baryon number 
and $L$ is the net lepton number of the Universe. 

Leptogenesis has drawn a significant attention as it demands lepton number violation with the hope to have possible 
connection with other lepton number violating processes. The canonical seesaw mechanism \cite{seesaw}, one of the most 
promising ways of explaining the origin of non-zero neutrino mass, also asks for lepton number violation via heavy neutral
singlet fermion exchange, i.e., the right-handed neutrino. The decays of these heavy particles can generate enough 
CP-asymmetry for successful leptogenesis. Thus neutrino mass generation and leptogenesis can be compatible with each 
other in this case. Another model of neutrino mass generation where leptogenesis occurs naturally is the Higgs triplet model. 
This is, of course, another realization of seesaw mechanism where a heavy Higgs scalar triplet is exchanged \cite{type-II}. 
One should notice that in both these cases lepton-number violation (by two units) occurs at a much higher scale than the 
scale of electroweak symmetry breaking ($\sim 10^2$ GeV). Note also that in the supersymmetric version of seesaw mechanism 
R-parity is conserved.  

On the other hand, R-parity violating models provide a source of neutrino masses and mixing, which is intrinsically 
supersymmetric in nature \cite{rpv-models}. In R-parity nonconserving SUSY, induced by lepton number violation by odd units,
realistic neutrino mass patterns and mixing angles can be generated compatible with the neutrino oscillation and reactor data. 
However, in the presence of these lepton number violating interactions at the scale of $10^3$ GeV or so, any pre-existing 
lepton asymmetry of the Universe would certainly be erased \cite{erasing_lepton_asymmetry}.

It was shown in \cite{hsm1,hsm2} that successful leptogenesis is possible in SUSY models with R-parity violation,
provided certain nonholomorphic lepton number violating interactions are taken into account. In addition, the 
familiar R-parity nonconserving interactions must be suppressed at the same time.  In such a scenario enough 
CP-asymmetry can be produced in the decay of the lightest neutralino into a charged Higgs boson and a lepton 
and this suppressed decay can also satisfy the out-of-equilibrium condition leading to successful leptogenesis.
On the other hand, the heavier neutralino must not satisfy the out-of-equilibrium condition and decays very fast.
It has also been shown that for a leptogenesis mechanism to be successful in the MSSM with R-parity violation, one 
must use only those lepton number violating terms, which are not constrained by neutrino masses. In this case the 
smallness of neutrino masses are explained by a radiative two-loop mechanism involving sneutrino-antisneutrino
mass splitting \cite{hsm2}. 

It has been shown in Ref.\cite{hsm2} that one needs a very heavy spectrum of SUSY particles to get correct values 
of the lepton asymmetry. In fact, a hierarchical scenario has been considered with the assumption that the bino-dominated
${\tilde \chi}^0_2$ is heavier than the wino-dominated ${\tilde \chi}^0_1$, i.e., $M_1 > M_2$. Here $M_1$ and $M_2$ 
are the $U(1)$ and $SU(2)$ gaugino mass parameters, respectively. Successful leptogenesis required that the gaugino
masses must be in the range of 2-6 TeV and hence this scenario might have a very remote possibility to be tested at
the LHC or the future ILC. We note in passing that the scales involved in the canonical leptogenesis models are very
high and impossible to be tested directly at the present or upcoming high energy colliders. 

On the other hand, it was shown in \cite{Pilaftsis1} that in models where leptogenesis is driven by the decays of
the right-handed neutrino ($M_{N_i}$), the CP-asymmetry can be enhanced for two nearly degenerate right-handed neutrinos. 
This is because in the limit $M_{N_i}-M_{N_j} << M_{N_i}$ the self-energy diagram dominates and due to this 
{\it resonance} effect the mass scale of the right-handed neutrinos can be lowered significantly. It was noted that 
sufficient lepton asymmetry can be generated even with $M_N ~ \sim$ 1 TeV \cite{Pilaftsis2}. The presence of TeV scale
right-handed neutrinos makes this scenario phenomenologically interesting compared to other canonical leptogenesis
scenarios. With this motivation, we revisited the model considered in Refs.\cite{hsm1,hsm2} and focused on a scenario
where we have nearly degenerate ${\tilde \chi}^0_2$ and ${\tilde \chi}^0_1$. We found that enough lepton asymmetry can be 
generated in this case through {\it Resonant Leptogenesis}, which can be converted into baryon asymmetry through 
the sphaleron processes. In our case the masses of ${\tilde \chi}^0_2$ and ${\tilde \chi}^0_1$ are found to be around
1 TeV and thus they have a possibility to be produced at the LHC. In addition, the lighter chargino is nearly degenerate
with the lighter neutralinos and can have a very slow decay leading to heavily ionizing charged tracks in the collider
detector. This could be a crucial test of the present model trying to explain leptogenesis and hence the baryon asymmetry 
of the Universe.  


\section*{Soft supersymmetry breaking and R-parity violation with nonholomorphic terms}
In a supersymmetric theory neither gauge invariance nor supersymmetry requires the conservation of lepton and baryon 
number. However, the lepton and baryon number violating operators can induce fast rate of proton decay and violate its
present experimental bound. To avoid this calamity a discrete symmetry called R-parity was introduced, which is defined as
\begin{equation}
{\rm R} \equiv (-1)^{3B+L+2S},
\end{equation}
where $B$ is the baryon number, $L$ the lepton number, and $S$ the spin 
angular momentum.  It is easy to check that the standard model particles 
have R = +1 and their supersymmetric partners have R = $-1$. An immediate consequence of R-parity conservation is that 
the lightest supersymmetric particle (LSP) is stable. On the other hand one notices that proton decay is still forbidden 
if either baryon number or lepton number is conserved in nature and R-parity is violated. This has led to considerable 
theoretical and phenomenological interest in studying models in which R-parity is violated. 

In an R-parity violating model, the superpotential can be written as 
\bea
W = W_{\rm MSSM} + W_{\rm RPV}.
\eea 
Here $W_{\rm MSSM}$ is the superpotential of R-parity conserving minimal supersymmetric standard model (MSSM) and is
given by
\bea
W_{\rm MSSM} = \mu H_1 H_2 + f_{ij}^e H_1 L_i e^c_j + f^d_{ij} H_1 Q_i d^c_j
+ f^u_{ij} H_2 Q_i u^c_j,
\label{superpot}
\eea
whereas the R-parity violating part of the superpotential is given by
\bea
W_{\rm RPV} = \mu_i L_i H_2 + \lambda_{ijk} L_i L_j e^c_k + \lambda'_{ijk}
L_i Q_j d^c_k + \lambda''_{ijk} u_i^c d_j^c d_k^c .
\label{rpv-superpot}
\eea

Here, $i,j,k$ are generational indices, $f^u_{ij}$, $f^d_{ij}$ and $f_{ij}^e$ are ($3 \times 3$) Yukawa matrices, 
$Q$, $u^c$, and $d^c$ are the quark doublet and singlet superfields and $L$ and $e^c$ are the lepton doublet and 
singlet superfields. The two Higgs doublet superfields are $H_1$ and $H_2$ giving rise to masses for the down-type quarks 
(and charged leptons) and the up-type quarks, respectively. The parameters $\mu$ and $\mu_i$ have
dimensions of mass and the terms $\mu_i L_i H_2$ are called the bilinear R-parity violating interactions whereas the
terms involving $\lambda$, $\lambda^\prime$ and $\lambda^{\prime\prime}$ are called trilinear R-parity violating
interactions.

Once supersymmetry is broken the soft supersymmetry breaking terms conserving 
R-parity and allowed by the standard model gauge group can be written as
\bea
{\cal L}_{soft}&=&
- \tilde{L}_i^{a \ast} (M_L^2)_{ij} \tilde{L}_j^a
- \tilde{e}_i^{c \ast} (M_e^2)_{ij} \tilde{e}_j^c
- \tilde{Q}_i^{a \ast} (M_Q^2)_{ij} \tilde{Q}_j^a
- \tilde{u}_i^{c \ast} (M_u^2)_{ij} \tilde{u}_j^c
\nonumber\\[1.5mm]
&&
- \tilde{d}_i^{c \ast} (M_d^2)_{ij} \tilde{d}_j^c
- M_{H_1}^2 H_1^{a \ast} H_1^a - M_{H_2}^2 H_2^{a \ast} H_2^a
- \varepsilon_{ab} ( B H_1^a H_2^b  +  h.c.)
\nonumber\\[1mm]
&&
-\varepsilon_{ab} \Big( (A_e f_e)_{ij} H_1^a \tilde{L}_i^b \tilde{e}_j^c
+ (A_u f_u)_{ij} H_2^b \tilde{Q}_i^a \tilde{u}_j^c
+(A_d f_d)_{ij} H_1^a \tilde{Q}_i^{b} \tilde{d}_j^c \, + \, h.c.
\Big)
\nonumber\\[1mm]
&&
-\frac{1}{2} \Big( M_3 \tilde{g} \tilde{g}
+ M_2 \tilde{W} \tilde{W}
+ M_1 \tilde{B} \tilde{B} \, + \, h.c. \Big)
\label{lsoft}\,.
\eea
Here, $a$ are $SU(2)$ indices. $M_3$, $M_2$, and $M_1$ are the $SU(3)$, $SU(2)$, and $U(1)$ gaugino mass parameters, 
respectively. $A_e$, $A_d$, and $A_u$ are the trilinear scalar couplings and $B$ is the Higgs bilinear parameter. 
The Higgs doublets giving mass to the standard model fermions are
\bea
H_1 = \left(\begin{array}{c}
h^0_1 
\\ h^-_1 
\end{array}\right), ~~~~~~~~~~~
H_2 = \left(\begin{array}{c}
h^+_2 
\\ h^0_2 \end{array}\right).
\eea
 
In an R-parity violating theory, additional soft terms may be present, and can be written as
\begin{equation}
{\cal L}_{soft}^{R \! \! \! /}=  - \varepsilon_{ab} \big( 
{B}_i' \tilde{L}_i^a H_2^b
+ {A}^{\prime e}_{ijk} \tilde{L}_i^a \tilde{L}_j^b \tilde{e}_k^c
+ {A}^{\prime d}_{ijk} \tilde{Q}_i^a \tilde{L}_j^b \tilde{d}_k^c \big)
- {A}^{\prime {\scriptstyle S}}_{ijk} \tilde{u}_i^c \tilde{d}_j^c \tilde{d}_k^c
\, + \, h.c. 
\label{lsoftRviol}\,.
\end{equation}
Following the convention of Ref.\cite{hsm2}, the coupling constants of all the 
R-parity conserving soft terms are denoted without a prime, while the R-parity
violating terms are denoted with a prime.

In principle there could be nonholomorphic terms in the Lagrangian. These nonholomorphic terms 
appear in the Lagrangian of the visible sector as an artifact of SUSY breaking in the hidden sector 
\cite{Hall-Randall,jack-jones,hetherington,sabanci}. 

The most general set of nonholomorphic soft terms conserving R-parity is: 
\begin{equation}
{\cal L}_{soft}^{NH}=
- N^e_{ij} H_2^{a \ast} \tilde{L}_i^a \tilde{e}_j^c
- N^d_{ij} H_2^{a \ast} \tilde{Q}_i^a \tilde{d}_j^c
- N^u_{ij} H_1^{a \ast} \tilde{Q}_i^a \tilde{u}_j^c \, + \, h.c.
\label{lsoftNH}\,.
\end{equation}
Similarly, nonholomorphic soft terms breaking R-parity are:
\begin{eqnarray}
{\cal L}_{soft}^{NH R \! \! \! /}&=&  
-{N}^{\prime {\scriptstyle B}}_i H_1^{a \ast} \tilde{L}_i^a
- {N}^{\prime e}_{i} H_2^{a \ast} H_1^a \tilde{e}_i^c
- {N}^{\prime u}_{ijk} \tilde{L}_i^{a \ast} \tilde{Q}_j^a \tilde{u}_k^c
\nonumber\\[1.5mm] 
&& 
- {N}^{\prime {\scriptstyle S}}_{ijk} \tilde{u}_i^c \tilde{e}_j^c \tilde{d}_k^{c \ast} 
- {N}^{\prime d}_{ijk} \varepsilon_{ab} \tilde{Q}_i^{a} \tilde{Q}_j^b 
\tilde{d}_k^{c \ast} \, + \, h.c.
\label{lsoftNHRviol}
\end{eqnarray}
In this paper we shall assume lepton-number violation but baryon-number conservation in the Lagrangian. This implies 
that $\lambda^{\prime \prime }_{ijk}=A^{\prime {\scriptstyle S}}_{ijk}= N^{\prime d}_{ijk}=0$.

If lepton number is violated by the bilinear R-parity breaking interactions $\mu_i L_i H_2$, then this
induces mixing between the neutrinos with the MSSM neutralinos. In addition, the sneutrinos (${\tilde \nu}_i$) 
may all acquire non-zero vacuum expectation values (VEVs). In this case the neutralino mass matrix gets enhanced
to a ($7\times 7$) mass matrix and in the basis $\left[ ~\tilde B,~~ \tilde W_3,
~~ \tilde h_1^0, ~~\tilde h_2^0, ~~\nu_1, ~~ \nu_2, ~~\nu_3~ \right]$
is given by
\begin{equation}
{\cal M} = \left[ \begin{array} {c@{\quad}c@{\quad}c@{\quad}c@{
\quad}c@{\quad}c@{\quad}c} 
M_1 & 0 & -s r_Z v_1  & s r_Z v_2 & 
-s r_Z v_{\nu_1} & -s r_Z v_{\nu_2} & 
-s r_Z v_{\nu_3}\\ 
0 & M_2 & c r_Z v_1  & -c r_Z v_2 & 
c r_Z v_{\nu_1} & c r_Z v_{\nu_2}& 
c r_Z v_{\nu_3}\\ 
-s r_Z v_1 & c r_Z v_1 & 0 & -\mu & 0 &0 &0\\ 
s r_Z v_2 & -c r_Z v_2 & -\mu & 0 & -\mu_1 & 
-\mu_2 & -\mu_3\\ 
-s r_Z v_{\nu_1} & c r_Z v_{\nu_1} & 0 & -\mu_1 & 0 &0 &0\\ 
-s r_Z v_{\nu_2} & c r_Z v_{\nu_2} & 0 & -\mu_2 & 0 &0 &0\\ 
-s r_Z v_{\nu_3} & c r_Z v_{\nu_3} & 0 & -\mu_3 & 0 &0 &0 
\end{array} \right],
\end{equation}
where $s = \sin \theta_W$, $c = \cos \theta_W$, $r_Z=M_Z/v$, and $v_1$, 
$v_2$, $v_{\nu_i}$ are the VEVs of $h_1^0$, $h_2^0$, and $\tilde \nu_i$ 
respectively, with $v_1^2 +v_2^2 +v_{\nu}^2 =v^2 \simeq$ (246 GeV)$^2$ and 
$v_\nu^2= v_{\nu_1}^2 + v_{\nu_2}^2 +v_{\nu_3}^2$.  We also define the parameter
$\tan \beta = v_2 / (v_1^2 + v_\nu^2)^{1/2}$.

In order to understand how a non-zero neutrino mass arises at the tree level from the above ($7\times 7$) mass 
matrix, let us assume that $\mu$ is much larger compared to the other entries. This means that 
$\tilde h^0_{1,2}$ form a heavy Dirac particle of mass $\mu$ which mixes very little  with the other
physical fields.  Integrating out these heavy fields one can write down the reduced ($5\times 5$) matrix using seesaw formula in the basis $\left[ ~\tilde B,~~ \tilde W_3,~~\nu_1, ~~ \nu_2, ~~\nu_3~ \right]$ as
\begin{equation}
{\cal M} = \left[ \begin{array} {c@{\quad}c@{\quad}c@{\quad}c@{\quad}c} 
M_1 - s^2 \delta r & sc \delta r & -s \epsilon_1 & -s \epsilon_2 & -s \epsilon_3\\ 
sc \delta r & M_2 - c^2 \delta r &  c \epsilon_1 &  c \epsilon_2 &  c \epsilon_3 \\ 
-s \epsilon_1 & c \epsilon_1 & 0 & 0 & 0\\
-s \epsilon_2 & c \epsilon_2 & 0 & 0 & 0\\
-s \epsilon_3 & c \epsilon_3 & 0 & 0 & 0
\end{array} \right],
\label{neutralino55}
\end{equation}
where 
\begin{eqnarray}
\delta &=& 2 M_Z^2 \frac{v_1 v_2}{v^2} \frac{1}{ \mu} = 
\frac{M_Z^2 \sin 2 \beta}{\mu} 
\sqrt{1-\frac{v_{\nu}^2}{v^2 \cos^2 \beta}}, 
\label{delta}
\end{eqnarray}
\begin{eqnarray}
\epsilon_i &=& \frac{M_Z}{v} \Big( v_{\nu_i} -  
\frac{\mu_i}{\mu} v_1 \Big),
\end{eqnarray}
\begin{eqnarray}
r&=&(1+M_2 / \mu \sin 2 \beta)/(1-M_2^2 / \mu^2).
\end{eqnarray}
Here the quantity $r$ has been introduced as a correction factor for finite values of $M_2/\mu$.

Looking at Eq.(\ref{neutralino55}), one can see that only the combination $\nu_l \equiv (\epsilon_1\nu_1 +
\epsilon_2 \nu_2 + \epsilon_3 \nu_3)/\epsilon$, with $\epsilon^2=\epsilon_1^2 + \epsilon_2^2 + \epsilon_3^2$, 
mixes with the gauginos. The other two orthogonal combinations decouple from the neutralino
mass matrix. In this case, only the eigenstate
\begin{equation}
\nu'_l = \nu_l + {s \epsilon \over M_1} \tilde B - {c \epsilon \over 
M_2} \tilde W_3,
\end{equation}
gets a seesaw mass given by
\begin{equation}
m_{\nu'_l} = - \epsilon^2 \left( {s^2 \over M_1} + {c^2 \over M_2} \right),
\label{massnu}
\end{equation}
whereas the other two neutrinos remain massless. These massless neutrino states may get non-zero
contribution to their masses through one-loop radiative corrections \cite{one-loop-bilinear}.

The two neutral gauginos mix with the neutrino $\nu_l $ and form mass eigenstates 
given by
\begin{eqnarray}
\tilde{\chi}_2^0 &=& \tilde B + {sc \delta r \over M_1 - M_2} \tilde W_3 - {s \epsilon 
\over M_1} \nu_l, \\ \tilde{\chi}_1^0 &=& \tilde W_3 - {sc \delta r \over M_1 - 
M_2} \tilde B + {c \epsilon \over M_2} \nu_l. \label{W3state}
\end{eqnarray}

Because of this non-zero neutrino component of the physical states ${\tilde \chi}^0_2$ and
${\tilde \chi}^0_1$, they can decay to lepton number violating two body final states such 
as {\bf ${\tilde \chi}^0_1 \rightarrow l^{\mp} W^{\pm}$} \cite{sr-bm,bm-sr-fv}. In general the gaugino 
masses can be complex and this can induce CP violation in the neutralino sector. Hence a lepton asymmetry 
can be generated from the lepton number violating decays of the neutralinos. 
However, the asymmetry generated this way is much smaller than the required value of $\sim 10^{-10}$ \cite{hsm2}.
This is because the asymmetry has to be much less than $(\epsilon/M_{1,2})^2$ (see, Eq.(\ref{W3state})). The quantity
$(\epsilon/M_{1,2})^2$ is of order ${m_{\nu^\prime_l}}/M_{1,2}$ (see, Eq.(\ref{massnu})), and hence the asymmetry is
$< 5 \times 10^{-13}$ if ${m_{\nu^\prime_l}} < 0.05$ eV, and $M_{1,2} > 100$ GeV. In addition, the out-of-equilibrium
condition on the decay width of the lightest neutralino results in an upper bound on $(\epsilon/M_{1,2})^2$, which is
independent of ${m_{\nu^\prime_l}}$. This effect also makes the asymmetry to be very much less than $10^{-10}$. 
Even if one considers R-parity violating trilinear couplings $\lambda$ and $\lambda^\prime$, it is possible to 
show \cite{hsm2} that they are also not compatible with the successful generation of a lepton or baryon asymmetry 
of the Universe. 


\section*{CP-asymmetry and Resonant Leptogenesis from neutralino decays}
For a leptogenesis mechanism to be successful in the MSSM with R-parity violation, one needs to satisfy two 
requirements. First the lepton-number violating terms must not be constrained by neutrino masses. Second we 
must satisfy the out-of-equilibrium condition for the lightest neutralino in such a way that the asymmetry is 
not automatically suppressed. In the line of \cite{hsm2}, let us first assume that the bino ${\tilde B}$ is heavier 
than the wino ${\tilde W}_3$. Although we will look at the scenario where the mass difference is very small. Because 
of R-parity violation, left- and right-chiral charged sleptons mix with the charged Higgs boson. Now if one assumes 
that the left-chiral charged slepton has a negligible mixing with the charged Higgs boson, then the ${\tilde \chi}^0_1$ 
decay into $l^\mp h^\pm$ is suppressed as long as the wino-bino mixing is small. This can be achieved if $\mu \gg M_1, M_2$. 
Hence the heavier neutralino ${\tilde \chi}^0_2$ decays quickly and the lighter neutralino ${\tilde \chi}^0_1$ has a much 
slower decay. At temperatures well above $T = M_{\rm SUSY}$, there are fast lepton number and R-parity violating interactions, 
which will wash out any $L$ or $B$ asymmetry of the Universe in the presence of sphalerons. This will be the case even at 
temperatures around $M_1$ (bino mass), when ${\tilde \chi}^0_1$ interactions violate $L_i$ as well as 
($B - 3L_i$) for $i = ~e, ~\mu, ~\tau$. Let us consider here that all other supersymmetric particles are heavier than the 
neutralinos, so that at temperatures below $M_1$ we need to consider only the interactions of ${\tilde \chi}^0_2$ and 
${\tilde \chi}^0_1$. 

We start with the well-known interaction{\bf s} of $\tilde B$ with $l$ and 
$\tilde l_R$ given by \cite{hk}
\begin{equation}
-{e \sqrt 2 \over \cos \theta_W} \left[ \bar l \left( {1 - \gamma_5 \over 
2} \right) {\tilde B} \tilde l_R + {\rm H.c.} \right].
\label{interac1}
\end{equation}

We then allow $\tilde l_R$ to mix with $h^-$, and $\tilde B$ to mix with 
$\tilde W_3$, so that the interaction of the physical state $\tilde{\chi}_1^0$ 
of Eq.(\ref{W3state}) with $l$ and $h^\pm$ is given by
\begin{equation}
\left( {sc\xi \delta r \over M_1 - M_2} \right) 
\left( {e \sqrt 2 \over \cos \theta_W} 
\right) \left[ \bar l \left( {1 - \gamma_5 \over 2} \right) \tilde 
{\chi}_1^0 h^- + {\rm H.c.} \right],
\label{interac2}
\end{equation}
where $\xi$ represents the $\tilde l_R - h^-$ mixing because of nonholomorphic R-parity violation and is assumed to be 
real. In the absence of nonholomorphic terms it is very hard to generate a large right-handed slepton and 
charged Higgs mixing without generating a large left-handed slepton and charged Higgs mixing as well. In order to achieve 
this we assume that $B^{'}_i$ and $N_i^{'B}$ in Eqs.(\ref{lsoftRviol}) and (\ref{lsoftNHRviol}) are negligible, thus the 
left-handed slepton and charged Higgs do not mix heavily. The term $N_i^{'e}$ produces the mixing between $\tilde{l}_R$ and 
charged Higgs. $\xi$ is proportional to $N_i^{'e}$ and measures the strength of the nonholomorphic coupling.
However, the parameter $\delta$ of Eq.(\ref{delta}) is complex. The nontrivial CP phase in the above contributes negligibly 
to the neutron electric dipole moment because the magnitude of $\delta$ is very small \cite{hsm2}.  

\begin{figure}[htbp]
\begin{center}
\centerline{\epsfig{file=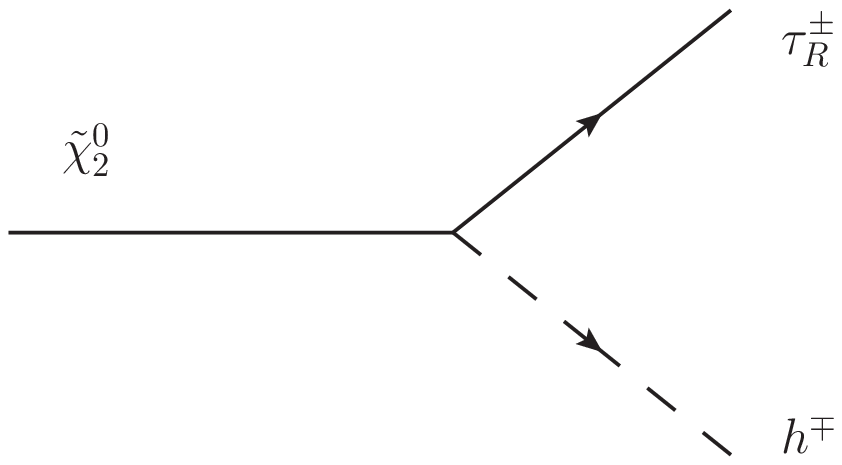,width=5.5 cm,height=3.5cm,angle=0}
\hskip 10pt \epsfig{file=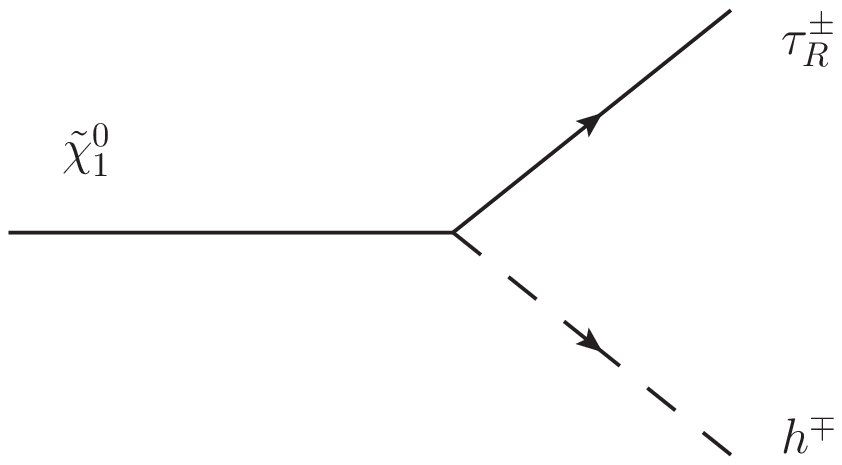,width=5.5 cm,height=3.5cm,angle=0}}
\vskip 10pt
\centerline{\epsfig{file=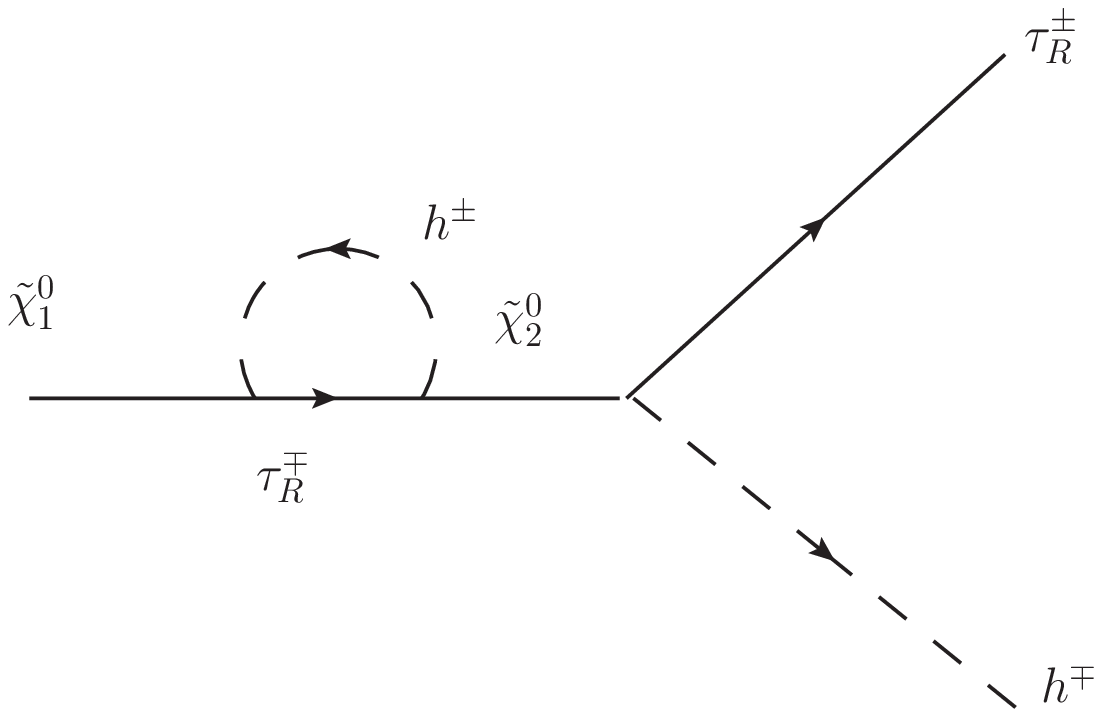,width=5.5 cm,height=4.5cm,angle=0}
\hskip 10pt \epsfig{file=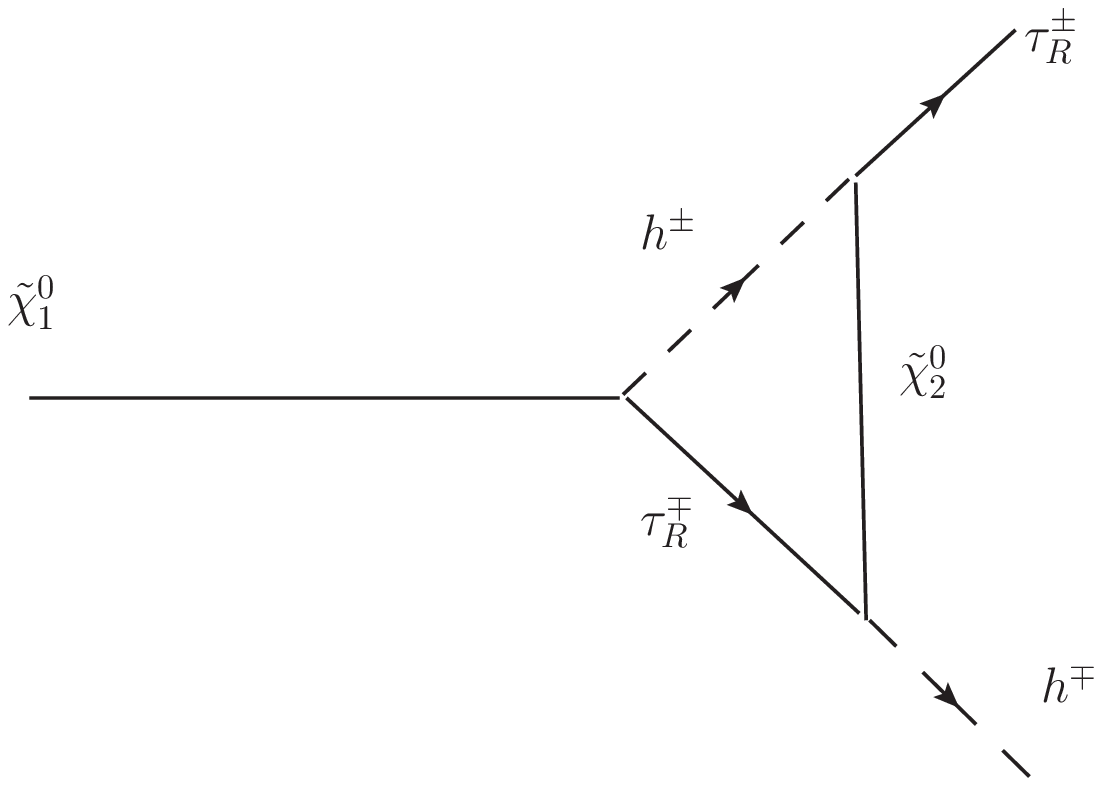,width=5.5 cm,height=4.5cm,angle=0}}
\caption{{\em Tree-level diagrams for (a) $\tilde{\chi}_2^0$ 
decay and (b) $\tilde{\chi}_1^0$ decay (through their Bino content), 
and the one-loop (c) self-energy and (d) vertex correction diagrams for  
$\tilde{\chi}_1^0$ decay.}} 
\label{figure1}
\end{center}
\end{figure}

In Fig.\ref{figure1} we show the lepton number violating decay processes 
(a) ${\tilde \chi}^0_2 \leftrightarrow l^\pm_R h^\mp$ and (b) ${\tilde \chi}^0_1 \leftrightarrow l^\pm_R h^\mp$,
at the tree level as well as the one-loop (c) self-energy and (d) vertex corrections of ${\tilde \chi}^0_1$ decay.
The decay width of $\tilde{\chi}_2^0$ is given by 
\begin{equation}
\Gamma_{\tilde{\chi}_2^0} = 
\Gamma ({\tilde{\chi}_2^0} \to {l^{+}} {h^{-}})+
\Gamma ({\tilde{\chi}_2^0} \to {l^{-}} {h^{+}})=
\frac{1}{4 \pi} \xi^2 
\frac{e^2}{c^2}
 {(M_{\tilde{\chi}_2^0}^2 - m_h^2)^2 \over 
M_{\tilde{\chi}_2^0}^3},
\end{equation}
while that of the $\tilde{\chi}_1^0$ is
\begin{equation}
\Gamma_{\tilde{\chi}_1^0} = 
\Gamma ({\tilde{\chi}_1^0} \to {l^{+}} {h^{-}})+
\Gamma ({\tilde{\chi}_1^0} \to {l^{-}} {h^{+}})
=\frac{1}{4 \pi} \xi^2 
\left( { e s |\delta| r \over M_1 - M_2} \right)^2 
{(M^2_{\tilde{\chi}_1^0} - m_h^2)^2 \over 
M_{\tilde{\chi}_1^0}^3}.\label{decW}
\end{equation}
Here $m_h$ is the mass of the charged Higgs boson ($h^-$). Let us also mention that for 
our choice of parameter (shown later) the radiative decay ${\tilde \chi}^0_2 
\rightarrow {\tilde \chi}^0_1 \gamma$ and the 3-body decay ${\tilde \chi}^0_2
\rightarrow {\tilde \chi}^0_1 f {\bar f}$ are very much suppressed and do not contribute
to the decay width. 

Calculating the interference between the tree-level and self-energy + vertex correction diagrams of 
Fig.\ref{figure1} one obtains the following CP-asymmetry from the decay of $\tilde{\chi}_1^0$ \cite{hsm2}:
\begin{eqnarray}
\in &=& \frac{\Gamma ({\tilde{\chi}_1^0} \to {l^{+}} {h^{-}})-
\Gamma ({\tilde{\chi}_1^0} \to {l^{-}} {h^{+}})}{\Gamma_{\tilde{\chi}_1^0}}\\ 
&=& {\alpha \xi^2 \over 2 \cos^2 \theta_W} 
{\mbox{Im} \delta^2 \over 
|\delta|^2} \left( 1 - {m_h^2 \over M_{\tilde{\chi}_1^0}^2} \right)^2 {x^{1/2} f(x) \over 
(1-x)}, \label{cp-asym}
\end{eqnarray}
where $x= M_{\tilde{\chi}_1^0}^2/M_{\tilde{\chi}_2^0}^2$ and
\begin{equation}
f(x) = 1 + {2(1-x) \over x} \left[ \left( {1+x \over x} \right) \ln (1+x) 
-1 \right].
\end{equation}

If the $\tilde{\chi}_1^0$ decay rate satisfies the out-of-equilibrium condition, then a lepton
asymmetry may be generated from the above decay asymmetry. As long as the lepton asymmetry is generated
at a temperature above, say $\sim$ 100 GeV, it will be converted to a baryon asymmetry of the 
Universe \cite{sphaleron_int}.

If the $\tilde{\chi}_1^0$ decay rate is much less than the expansion rate 
of the Universe, the generated lepton asymmetry is of the order of the decay 
asymmetry given in Eq.(\ref{cp-asym}). In other words, the out-of-equilibrium condition 
reads as:
\begin{equation}
K_{\tilde{\chi}_1^0} = {\Gamma_{\tilde{\chi}_1^0} \over
H(M_{\tilde{\chi}_1^0})} \ll 1 , 
\end{equation}
where $H(T)$ is the Hubble constant at the temperature $T$ and is given by
\begin{equation}
H(T)=\sqrt{\frac{4 \pi^3 g_*}{45}} \frac{T^2}{M_{Planck}},\label{hubble}
\end{equation}
with $g_*$ the number of massless degrees of freedom which is 106.75 in this case corresponding to the
standard model (SM) degrees of freedom and $M_{Pl} \sim 10^{19}$ GeV is the Planck scale \footnote{In our 
numerical analysis we have used $M_{Pl} = 0.9 \times 10^{19}$ GeV.}.

However, in order to to present a realistic and reliable estimation of the lepton asymmetry generated from 
neutralino decay we solve the full Boltzmann equations \cite{kt}. In our scenario we have the Boltzmann 
equations (including the wash-out effects) same as in \cite{hsm2}
\begin{eqnarray}
\frac{dX_{\tilde{\chi}_1^0}}{d z}&=&
- z K_{\tilde{\chi}_1^0} \frac{K_1(z)}{K_2(z)}   
\Big({X_{\tilde{\chi}_1^0}}-{X^{eq}_{\tilde{\chi}_1^0}}  \Big),\nonumber\\
\frac{d X_L}{d z}&=&
z K_{\tilde{\chi}_1^0}   \frac{K_1(z)}{K_2(z)} 
\Big[\varepsilon ({X_{\tilde{\chi}_1^0}}-{X^{eq}_{\tilde{\chi}_1^0}} )
- \frac{1}{2}   \frac{X_{\tilde{\chi}_1^0}}{X_\gamma} X_L \Big]\nonumber\\
&&- z  \Big(\frac{M_{\tilde{\chi}_2^0}}
{M_{\tilde{\chi}_1^0}} \Big)^2 
K_{\tilde{\chi}_2^0} \Big[\frac{1}{2} \frac{K_1(z M_{\tilde{\chi}_2^0}/M_{\tilde{\chi}_1^0})}
{K_2(z M_{\tilde{\chi}_2^0}/M_{\tilde{\chi}_1^0})}
\frac{X_{\tilde{\chi}_2^0}}{X_\gamma} X_L 
+2 \frac{X_L}{X_\gamma} 
\frac{\gamma^{eq}_{scatt.}}{\circledS 
\Gamma_{\tilde{\chi}_2^0} } \Big],\label{boltzmann}
\end{eqnarray}    
where $K_1$, and $K_2$ are the modified Bessel's functions, $z \equiv M_{\tilde{\chi}_1^0}/T$, 
$K_{\tilde{\chi}_2^0}=\Gamma_{\tilde{\chi}_2^0} / H(M_{\tilde{\chi}_2^0})$,
and $\circledS =g_* \frac{2 \pi^2}{45} T^{3}$ is the entropy density. The number densities per
comoving volume have been defined as $X_i = n_i/\circledS$ in terms of the number densities of particles `i'.

Once the wash-out effects are included (as in Eq.(\ref{boltzmann})) it is very hard to generate the lepton asymmetry 
of correct order ($O (10^{-10})$) keeping nearly degenerate neutralinos $\sim$ 1 TeV. In  order
to find a reliable and stable solution we need very large values of $\mu \sim 40-75$ TeV and $\tan \beta \sim 55-60$. 
We consider three sets of parameters to estimate the lepton asymmetry:
\bea
{\bf Set ~I:~}& M_{\tilde{\chi}_2^0}=1520.998~{\rm GeV},~M_{\tilde{\chi}_1^0}=1520.997~{\rm GeV},~ \tan \beta=64,~\xi = 1.85 \times 10^{-5},\nonumber \\
 &\mu = 75 ~{\rm TeV}, M_1 = 1521~{\rm GeV},~M_2 = 1520.3~{\rm GeV}, 
\label{Set I} 
\eea
\bea
{\bf Set~ II:~}& M_{\tilde{\chi}_2^0}=1380.9998~{\rm GeV},~M_{\tilde{\chi}_1^0}=1380.9997~{\rm GeV},~\tan \beta=72,~\xi = 0.64 \times 10^{-5},  \nonumber \\
 & \mu = 43~ {\rm TeV}, ~M_1 = 1381~{\rm GeV},~M_2 = 1380.4~{\rm GeV},
\label{Set II}
\eea
and
\bea
{\bf Set ~III:~}& M_{\tilde{\chi}_2^0}=1680.9~{\rm GeV},~M_{\tilde{\chi}_1^0}=1680.8~{\rm GeV},~ \tan \beta=65,~\xi = 0.67 \times 10^{-5},\nonumber \\
 &\mu = 43 ~{\rm TeV}, M_1 = 1681.0~{\rm GeV},~M_2 = 1680.4~{\rm GeV},
\label{Set III}
\eea
with $m_h=180~{\rm GeV},~M_Z=91.19~{\rm GeV}$.
\begin{figure}[htbp]
\begin{center}
{\epsfig{file=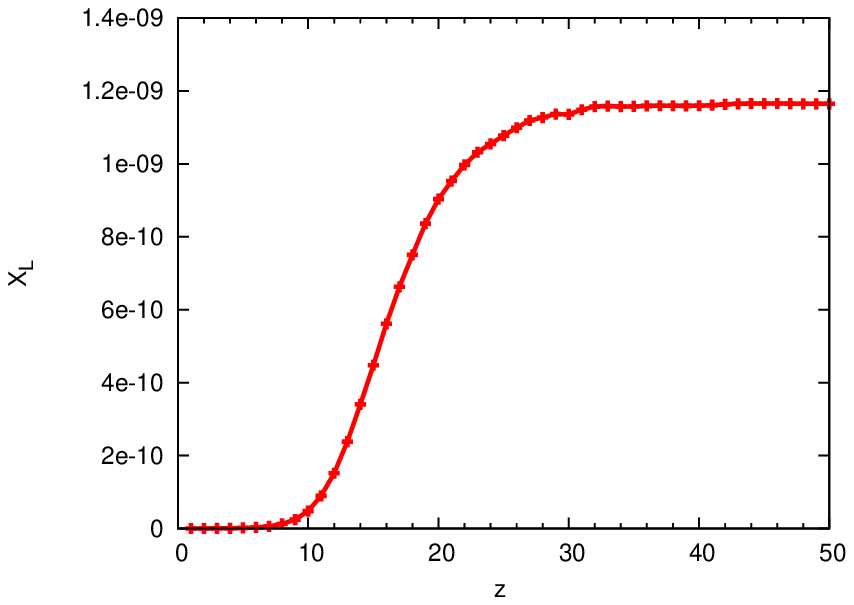,width=5. cm,height=6.3cm}}
\hskip 0.2cm
{\epsfig{file=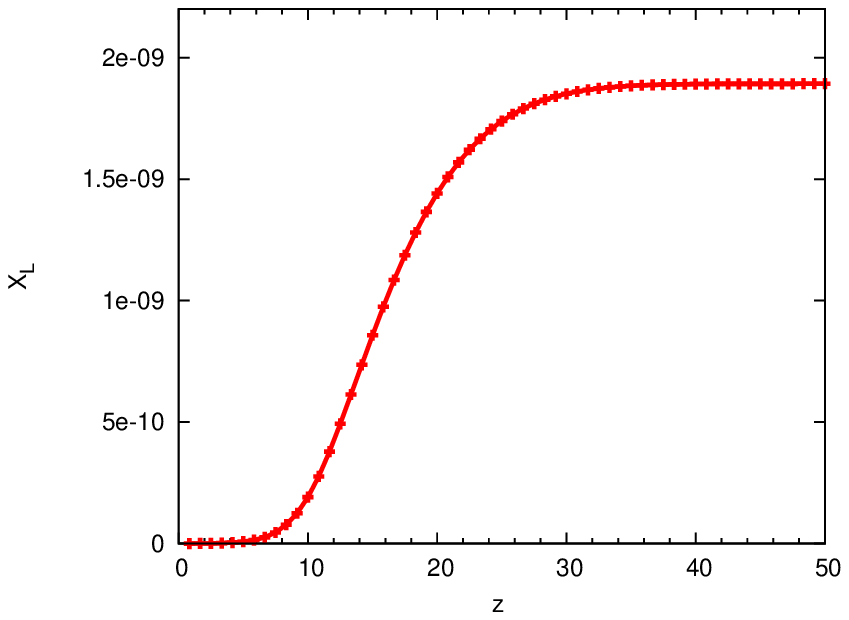,width=5. cm,height=6.cm}}
\hskip 0.2cm
{\epsfig{file=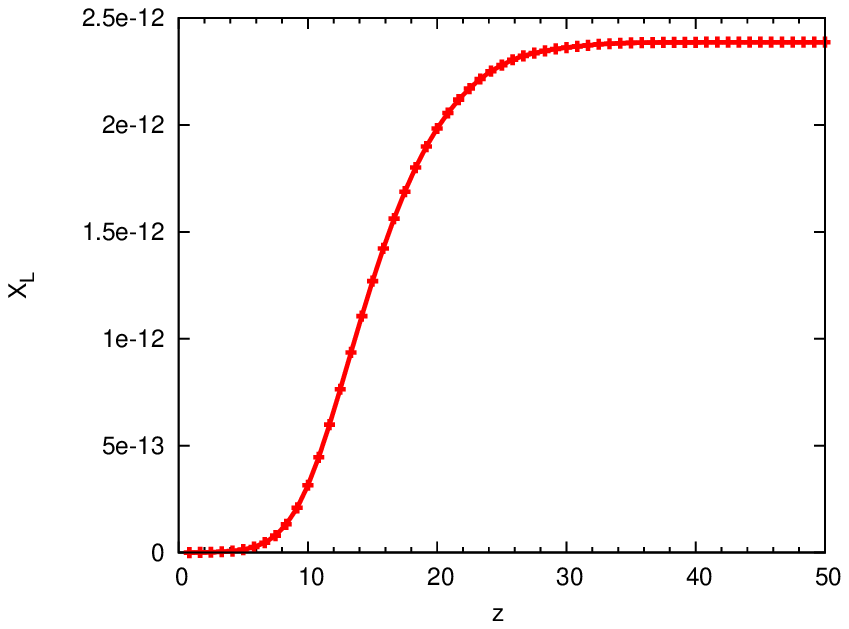,width=5. cm,height=6.cm}}
\caption{\em Lepton asymmetry ($X_L$) vs z with $g_*=106.75$ for Set I (left), Set II (middle), and Set III (right).} 
\end{center}
\label{fig1}
\end{figure}
\vskip 0.1cm
The resulting evolution of the lepton asymmetry ($X_L$) is shown in Fig.2\ref{fig1}(left and middle) for the
parameter choices of Set I and Set II, respectively. We see from these two figures that a large asymmetry of
order $10^{-10}$ is produced at $T \sim M_Z$ provided we have nearly degenerate ${\tilde \chi}^0_2$ and
${\tilde \chi}^0_1$, very large $\mu$ and $\tan\beta$ and values of $\xi$ around $1 \times 10^{-5}$. 
In Fig.2(right) the lepton asymmetry is shown for the parameters given in Set III. 
Note that in order to have a large lepton asymmetry the splitting between $M_{{\tilde \chi}^0_2}$ and 
$M_{{\tilde \chi}^0_1}$ is required to be much smaller compared to the splitting between the gaugino parameters 
$M_1$ and $M_2$. But when the splitting between $M_{{\tilde \chi}^0_2}$ and $M_{{\tilde \chi}^0_1}$ is increased 
a little bit, lepton asymmetry falls very sharply. This shows the importance of the requirement of a very highly 
degenerate neutralinos. On the other hand, this highly degenerate neutralino scenario (with larger splitting between 
$M_1$ and $M_2$) is difficult to achieve in practice and might need additional fine tuning in this model or could be 
taken as a hint in favour of non-minimal SUSY models. It is also important to note that the above discussion is based 
on the tree-level neutralino mass matrix. One should also include radiative corrections at one-loop order to check the 
stability of the results presented here. However, that is a subject of a separate study and we will not take it up in 
the present paper. Our main objective here is to present the idea of resonant leptogenesis in the MSSM with 
nonholomorphic R-parity violating soft SUSY breaking interactions and its testability at the LHC. 

\section*{Large values of $\tan \beta$ and $\mu$}
In the general MSSM scenario the maximum value of $\tan \beta \lesssim 50$ is restricted by the perturbative limit of 
the supersymmetric Yukawa couplings. Nevertheless, the possibility of large values of $\tan\beta (\gg 50)$ has been 
considered in various context within the MSSM. For example, in the context of up-down Yukawa unification and Higgs 
mediated FCNC, $\tan\beta \gg 50$ has been pointed out \cite{hamzaoui-pospelov,hamzaoui-pospelov-toharia}. Some other
aspects of Higgs phenomenology with $\tan\beta$ as large as 130 have been studied in \cite{carena-heinemeyer}. Very 
recently the authors of \cite{bogdan-fox} have pointed out that the parameter space of MSSM includes a region where 
the down-type fermion masses are generated by the loop induced couplings to the up-type Higgs doublet. In this region of
MSSM, a large value of $\tan\beta \gsim 100$ is consistent with the perturbativity of the SUSY Yukawa couplings of 
the down-type fermions to $H_1$.   
 \begin{figure}[htbp]
 \vspace*{-2cm}
 \begin{center}
 {\epsfig{file=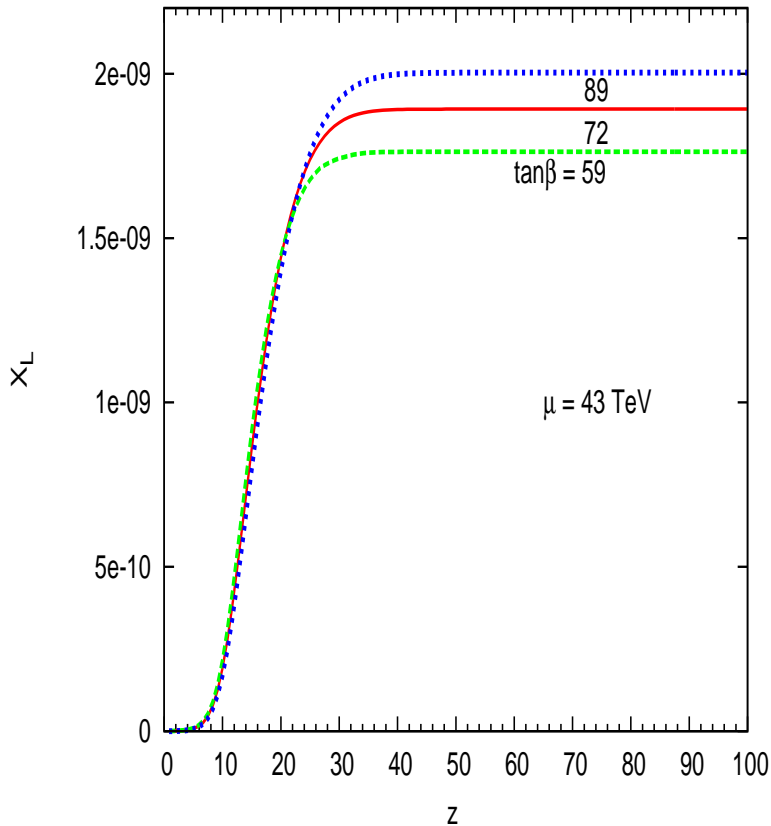,width=10.5 cm,height=11.0cm}}
 \vspace*{0.5cm}
 \caption{\em Lepton asymmetry vs $\tan \beta$ with $g_*=106.75$ for Set II.} 
 \end{center}
 \label{fig3}
 \end{figure}
In Fig.3\ref{fig3} we have shown the variation of lepton asymmetry with $\tan \beta$, keeping $\tan \beta$ 
large. From the above plot it is clear that as we increase $\tan \beta$ the lepton asymmetry 
increases keeping other parameters fixed. This justifies our choice of large value for $\tan \beta$.


The bilinear R-parity conserving term $\mu H_1 H_2$ in Eq.(\ref{superpot}) introduces $\mu$ as a free parameter of the theory. 
There is no known symmetry that protects $\mu$ from having a value $\sim M_{Pl}$. However, from phenomenological point of view 
one would expect that the value of $\mu$ should be around 100 GeV or 1 TeV scale to avoid unnatural fine tuning in the theory. 
This is evident from the electroweak symmetry breaking condition that connects $\mu$ with the mass of the Z-boson by the 
following relation (in the limit of large $\tan\beta$) \cite{martin-primer}
\begin{eqnarray}
m^2_Z = -2(|\mu|^2+m_{H_2}^2) + \frac{2}{\tan^2\beta}(m^2_{H_1} - m^2_{H_2}) + {\cal O}(1/{\tan^4\beta}).
\label{ewsb-relation}
\end{eqnarray}
In order to have the correct value for $m_Z$, the input parameters $m_{H_2}^2$, $m^2_{H_1}$, and $\mu$ on the right hand side 
of Eq.(\ref{ewsb-relation}) should be within an order of magnitude or two of $m^2_Z$ in the absence of any fine cancellation 
between various terms. However, if one admits some amount of fine tuning then the value of $\mu$ in the range of 40-45 TeV 
is possible. It must be noted here that typical viable solutions for the MSSM still requires significant 
cancellation \cite{martin-primer}. We have seen in this paper that to achieve a reliable and stable solution for the lepton 
asymmetry, $\mu$ needs to be very large, i.e., in the range of 40-75 TeV. Thus this situation might be more fine-tuned than 
the general MSSM scenario. We note in passing that in some studies \cite{kanemura-matsuda-ota} in the context of Lepton Flavour 
Violation (LFV) in the Higgs boson decay large values of $\mu$ ($\sim$ 25 TeV) has been suggested.

The results from the studies cited above can also be applied in the scenario under consideration. For example,
Ref. \cite{carena-heinemeyer} pointed out that $\mu$ has to be large and positive and $\tan\beta >$ 50.
Thus in our case we find solutions for the lepton asymmetry for low bino and wino masses ($\sim$ 1 TeV)
while $\mu$ is very large $\sim$ 40 TeV and $\tan\beta >$ 50. Having rather low $M_{\tilde{\chi}_1^0}$, $M_{{\tilde \chi}^\pm_1}$  
and $M_{\tilde{\chi}_2^0}$ can give us possible distinct and interesting signatures at the LHC and thus
this model becomes testable in the near future. 

Because of large $\mu$ higgsino sector is decoupled from wino and bino sector. The third and the fourth 
neutralino as well as the heavier chargino are very massive (as their masses are controlled mostly by $\mu$) 
and are out of reach of the LHC. On the other hand, the lightest and the next-to-lightest neutralino and the lighter 
chargino are within the reach of the LHC. 
In addition since the lighter chargino is also nearly degenerate with the 
lighter neutralinos, it is expected that the lighter chargino will produce heavily ionizing charged tracks in the detectors
at the LHC. It must be noted though that the pair production cross sections of a 1 TeV ${\tilde \chi}^0_1$ or
${\tilde \chi}^\pm_1$ is of the order of $10^{-2}$ fb at the LHC with $\sqrt{s} = 14$ TeV. This means one can see 
signals of such a scenario only with a large integrated luminosity ($\sim$ 300 fb$^{-1}$). 
Nevertheless, from the above discussion it is clear that this model of leptogenesis is phenomenologically rich and 
possibly testable at the LHC. In \cite{bcr} a class of high scale nonuniversal scenario is suggested where at the 
electroweak scale nearly degenerate neutralinos can be achieved. 
It is noted in \cite{hsm2} that the nonholomorphic terms $N_1^{'e}$ can generate 
neutrino masses. 
The light neutrino mass is given as:
\begin{eqnarray}
m_\nu&=&\frac{1}{256 \pi^4} 
\frac{e^2}{\sin^2 \theta_W} \mu^2
\frac{m_{\tau}^2}{v^2} \xi^2  M_{\tilde{\chi}_1^0}
\frac{
 M^2_{\tilde \nu} -M_{\tilde{\chi}_1^0}^2- M_{\tilde{\chi}_1^0}^2 \ln \Big( 
 M^2_{\tilde \nu}  / M_{\tilde{\chi}_1^0}^2 \Big) }{\Big(
 M^2_{\tilde \nu} - M_{\tilde{\chi}_1^0}^2 \Big)^2},
\end{eqnarray}
when the slepton ${\tilde l}^+$ that mixes with $h^+$ is mainly ${\tilde \tau}^+$. 
In our scenario, we find that correct order of neutrino mass (0.024 eV) is achieved for the 
parameters in Set II (see Eq.(\ref{Set II})) with $M_{\tilde \nu}$ = 1381.2 GeV. 
For the other set of parameters correct order of neutrino masses can be generated with suitable choice of 
$M_{\tilde \nu}$. Thus a realistic scheme of radiative neutrino mass generation which originates from the same 
nonholomorphic terms can be accommodated along with other phenomenological aspects, discussed in earlier sections, 
in our scenario.
\vskip 20pt

\section*{Conclusions}
We discuss the possibility of resonant leptogenesis in an R-parity violating supersymmetric
standard model with nonholomorphic supersymmetry breaking terms in the scalar potential.
We work within a parameter space where the lighter neutralinos, $\tilde{\chi}_1^0$ and  $\tilde{\chi}_2^0$ 
are nearly degenerate in mass. In this framework we find out a consistent scenario
where neutrino masses, low scale leptogenesis, and interesting collider signatures exist simultaneously. As the 
masses of the $\tilde{\chi}_1^0$, $\tilde{\chi}_2^0$ and ${\tilde \chi}^\pm_1$ are around the TeV scale, they can 
be produced at the LHC with small but finite cross-sections. This makes our model phenomenologically rich and 
accessible at the LHC with signatures involving heavily ionizing charged tracks. Thus this low scale resonant 
leptogenesis model is testable at the LHC which might give an indication of the presence of nonholomorphic couplings. 

\vskip 10pt

{\bf Acknowledgements:} JC wishes to thank Ashoke Sen for helpful discussions. 
JC would like to thank the Department of Theoretical Physics, Indian Association 
for the Cultivation of Science (IACS) for the hospitality where a part of this work was done. 
This research has been supported by funds from the XIth Plan `Neutrino physics' and 
RECAPP projects at the Harish-Chandra Research Institute (HRI), Allahabad.


\begin{thebibliography}{99}
\bibitem{nakamura-pdg}
K. Nakamura {\it et al.}, J. Phys. G {\bf 37}, 075021 (2010).

\bibitem{leptogenesis}
For a recent review, see, for example, Mu-Chun Chen, hep-ph/0703087 and references therein.

\bibitem{seesaw}
  P.~Minkowski,
Phys. Lett. B {\bf 67}, 421 (1977);
M.~Gell-Mann, P.~Ramond and R.~Slansky,
{\it {Complex Spinors And Unified Theories}},
Published in Supergravity, P. van Nieuwenhuizen $\&$ D.Z. Freedman (eds.),
North Holland Publ. Co., 1979. Published in Stony Brook Wkshp.1979:0315 (QC178:S8:1979).
T.~Yanagida,
{\it{in \emph{Proceedings of the Workshop on the Unified Theory
and the Baryon Number in the Universe} (O.~Sawada and A.~Sugamoto, eds.), KEK,
  Tsukuba, Japan, 1979, p.~95}};
S.~L. Glashow,
{\it{\emph{The future of elementary particle physics}, in
  \emph{Proceedings of the 1979 Carg{\`e}se Summer Institute on Quarks and
  Leptons} (M.~L{\'e}vy, J.-L. Basdevant, D.~Speiser, J.~Weyers,
R.~Gastmans, and M.~Jacob, eds.), Plenum Press, New York, 1980, pp.~687--713}};
R.~N.~Mohapatra and G.~Senjanovic,
Phys. Rev. Lett {\bf 44}, 912 (1980).

\bibitem{type-II}
M.~Magg and C.~Wetterich,
  Phys.\ Lett.\  B {\bf 94}, 61 (1980);
G.~Lazarides, Q.~Shafi and C.~Wetterich,
  Nucl.\ Phys.\  B {\bf 181}, 287 (1981);
R.~N.~Mohapatra and G.~Senjanovic,
  Phys.\ Rev.\  D {\bf 23}, 165 (1981);
J.~Schechter and J.~W.~F.~Valle,
  Phys.\ Rev.\  D {\bf 25}, 774 (1982);
J.~Schechter and J.~W.~F.~Valle,
  Phys.\ Rev.\  D {\bf 22}, 2227 (1980).


\bibitem{rpv-models}
For a review, see, for example, 
R.~Barbier {\it et al.},
Phys. Rept. {\bf 420}, 1 (2005)[hep-ph/0406039];\\
M.~Chemtob,
  Prog. Part. Nucl. Phys. B {\bf 54}, 71 (2005)[hep-ph/0406029].

\bibitem{erasing_lepton_asymmetry}
B.A. Campbell, S. Davidson, J.E. Ellis, and K. Olive, Phys.\ Lett.\ B {\bf 256}, 457 (1991); 
W. Fischler, G.F. Giudice, R.G. Leigh, and S. Paban, Phys.\ Lett.\ B {\bf 258}, 45 (1991);
H. Dreiner and G.G. Ross, Nucl.\ Phys.\ B {\bf 410}, 188 (1993); J.M. Cline, K. Kainulainen, and K.A.
Olive, Phys.\ Rev.\ D {\bf 49}, 6394 (1994); E. Ma, M. Raidal, and U. Sarkar, Phys.\ Lett.\ B {\bf 460}, 359 
(1999).  
\bibitem{hsm1}
  T.~Hambye, E.~Ma and U.~Sarkar,
  Phys.\ Rev.\  D {\bf 62}, 015010 (2000)
  [arXiv:hep-ph/9911422].


\bibitem{hsm2}
  T.~Hambye, E.~Ma and U.~Sarkar,
  Nucl.\ Phys.\  B {\bf 590}, 429 (2000)
  [arXiv:hep-ph/0006173].


\bibitem{Pilaftsis1}
  A.~Pilaftsis,
  Phys.\ Rev.\  D {\bf 56}, 5431 (1997)
  [arXiv:hep-ph/9707235].

\bibitem{Pilaftsis2}
  A.~Pilaftsis and T.~E.~J.~Underwood,
  Nucl.\ Phys.\  B {\bf 692}, 303 (2004)
  [arXiv:hep-ph/0309342].


\bibitem{Hall-Randall}
  L.~J.~Hall and L.~Randall,
  Phys.\ Rev.\ Lett.\  {\bf 65}, 2939 (1990).

\bibitem{jack-jones}
I. Jack and D.R.T. Jones, Phys.\ Rev.\ D {\bf 61}, 095002 (2000).

\bibitem{hetherington}
J.P.J. Hetherington, J. High Energy Phys. {\bf 10} (2001) 024.

\bibitem{sabanci}
A. Sabanci, A. Hayreter, and L. Solmaz, Phys.\ Lett.\ B {\bf 661}, 154 (2008).

\bibitem{one-loop-bilinear}
See, e.g., Y. Grossman and H.E. Haber, Phys.\ Rev.\ D {\bf 59}, 093008 (1999); 
E.J. Chun and S.K. Kang, Phys.\ Rev.\ D {\bf 61}, 075012 (2000); M. Hirsch, M.A.
Diaz, W. Porod, J.C. Romao and J.W.F. Valle, Phys.\ Rev.\ D {\bf 62}, 113008 (2000);
M.A. Diaz, M. Hirsch, W. Porod, J.C. Romao and J.W.F. Valle, Phys.\ Rev.\ D {\bf 68},  013009 
(2003) [Erratum {\it ibid.} D {\bf 71}, 059904 (2005)]; S. Davidson and 
M. Losada, J. High Energy Phys. {\bf 05} (2000) 021; Y. Grossman and S. Rakshit,
Phys.\ Rev.\ D {\bf 69},  093002 (2004); A. Dedes, S. Rimmer and J. Rosiek, 
J. High Energy Phys. {\bf 08} (2006) 005; 
R.~S.~Hundi, arXiv:1101.2810 [hep-ph].
 

\bibitem{sr-bm}
S. Roy and B. Mukhopadhyaya, Phys.\ Rev.\ D {\bf 55}, 7020 (1997). 

\bibitem{bm-sr-fv}
B. Mukhopadhyaya, S. Roy, and F. Vissani, Phys.\ Lett.\ B {\bf 443}, 191 (1998).

\bibitem{hk} 
H. E. Haber and G. L. Kane, Phys.\ Rep.\ {\bf 117}, 75 (1985).

\bibitem{sphaleron_int}
V.A. Rubakov and M.E. Shaposhnikov, Usp. Fiz. Nauk {\bf 166}, 493 (1996); Phys. Usp. {\bf 39}, 461 (1996);
A. Riotto, hep-ph/9807454; G.D. Moore, Phys.\ Rev.\ D {\bf 59}, 01450 (1999)3.




\bibitem{kt}
  E. W. Kolb and M. S. Turner, {\em The Early Universe} (Addison-Wesley, 
Reading, MA, 1990); J.N. Fry, K.A. Olive and M.S. Turner, Phys. Rev. Lett. 
{\bf 45}, 2074 (1980); Phys.\ Rev.\ D {\bf 22}, 2953 (1980); Phys.\ Rev.\ D {\bf 22}, 2977 
(1980); E.W. Kolb and S. Wolfram, Nucl.\ Phys.\ B {\bf 172}, 224 (1980).

\bibitem{hamzaoui-pospelov}
C. Hamzaoui, M. Pospelov, Eur. Phys. J. C {\bf 8} 151 (1999).

\bibitem{hamzaoui-pospelov-toharia}
C. Hamzaoui, M. Pospelov, M. Toharia, 
Phys.\ Rev.\ D {\bf 59}, 095005 (1999).

\bibitem{carena-heinemeyer}
M.S. Carena, S. Heinemeyer, C.E.M. Wagner, G. Weiglein, 
Eur. Phys. J. C {\bf 45} 797 (2006).

\bibitem{bogdan-fox}
  B.~A.~Dobrescu and P.~J.~Fox,
  Eur. Phys. J. C {\bf 70} 263 (2010) [arXiv:1001.3147/hep-ph].

\bibitem{martin-primer}
S.P. Martin, arXiv:hep-ph/9709356.

\bibitem{kanemura-matsuda-ota}
  S.~Kanemura, K.~Matsuda, T.~Ota, T.~Shindou, E.~Takasugi and K.~Tsumura,
  Phys.\ Lett.\  B {\bf 599}, 83 (2004)
  [arXiv:hep-ph/0406316]; 
 S.~Kanemura, K.~Matsuda, T.~Ota, T.~Shindou, E.~Takasugi and K.~Tsumura,
  arXiv:hep-ph/0408276.

\bibitem{bcr}
  S.~Biswas, J.~Chakrabortty and S.~Roy,
Phys.\ Rev.\  D {\bf 83}, 075009 (2011) 
  [arXiv:1010.0949/hep-ph].






\end{thebibliography}
\end{document}